\documentclass[prd,twocolumn,tightenlines,showpacs,nofootinbib,amsfonts,amssymb,amsmath]{revtex4}
\usepackage{graphicx,epsfig,rotate}
\usepackage{bm}           

\newcommand{\dd}{{\rm d}}
\newcommand{\bx}{{\bm{x}}}
\newcommand{\bk}{{\bm{k}}}
\newcommand{\Dirac}{\delta_{\rm D}}
\newcommand{\obs}{{_O}}
\newcommand{\obsup}{{^O}}
\newcommand{\emit}{{_E}}
\newcommand{\mat}{{_{\rm m}}}
\begin{document}

\title{Time drift of cosmological redshifts and its variance}

\author{Jean-Philippe Uzan}
 \email{uzan@iap.fr}
 \affiliation{
             Institut d'Astrophysique de Paris,
             Universit\'e Pierre~\&~Marie Curie - Paris VI,
             CNRS-UMR 7095, 98 bis, Bd Arago, 75014 Paris, France,}

\author{Francis Bernardeau}
 \email{fbernard@spht.saclay.cea.fr}
 \affiliation{Service de Physique Th{\'e}orique,
         CEA/DSM/SPhT, Unit{\'e} de recherche associ{\'e}e au CNRS, CEA/Saclay
         91191 Gif-sur-Yvette c{\'e}dex, France}

\author{Yannick Mellier}
\email{mellier@iap.fr}
 \affiliation{
             Institut d'Astrophysique  de Paris,
             Universit\'e Pierre~\&~Marie Curie - Paris VI,
             CNRS-UMR 7095, 98 bis, \ Bd Arago, 75014 Paris, France,}

\begin{abstract}
The contribution of cosmological perturbations to the time drift
of the cosmological redshift is derived. It is shown that the
dominant correction arises from the local acceleration of both the
emitter and the observer. The amplitude of this effect is
estimated to be of the order of 1\% of the drift signal at
$z=2-4$, but can easily be lowered down to 0.1\% by using many
absorption lines and quasars.
\end{abstract}
 \date{12 November 2007}
 \maketitle

\section{Introduction}

The increasing observational evidences that the expansion of the
universe is accelerating (see Ref.~\cite{refsn,refdata} and
reference therein) has stimulated a rising interest to the
reconstruction of its expansion history. An important outcome of
these theoretical studies is to clarify the sensitivity of
observational tests to dark energy properties and to assess how
each could be corrupted by extra-noise from other cosmological
effects. Comprehensive investigations of these nuisances have been
carried out on ``standard" tests, like SNIa, weak lensing, BAO,
CMB, ISW or clusters of galaxies.

In contrast, very few focussed on the time drift effect that
changes the observed redshift  of an object as function of time.
The recent claim that it  may drive the conceptual design of
instrument for next generation giant telescopes raised the need
that similar attention should be paid to the theoretical ground of
this novel technique. Interestingly, such an observation may lead
to a better understanding of the physical origin of the recent
acceleration~\cite{refde,refde2} and to a determination of the
dark energy equation of states~\cite{Lake07} as well as
constraints on dark energy models~\cite{demodels} or tests of the
variation of fundamental constants~\cite{constant,Molaro}.

As first pointed out by Sandage~\cite{Sandage}, in a homogeneous
and isotropic spacetime, the time drift of the observed redshift
is directly related to the Hubble function by
\begin{equation}\label{SLformula}
 \dot z = (1+z)H_0-H(z)\equiv \dot{\bar z}(\eta_0,z)\ .
\end{equation}
Given the  most likely ranges of cosmological parameter values
derived from observations, in a $\Lambda$CDM model the typical
amplitude of the redshift drift is of order $\delta z\sim
-4\times10^{-10}$ on a  time scale of $\delta t= 10$~yr, for  a
source at redshift $z=4$. This corresponds to a spectroscopic tiny
velocity shift, $\delta v \equiv c\delta z/(1+z)$, of $\delta
v\sim 2.5\ {\rm cm/s}$. Fig.~\ref{AmpError} (left panel) shows the
time drift as function of redshift for the standard $\Lambda$CDM
model and a dark energy models with an equation of state changed
by only 10\% ($w=-0.9$), all other parameters being kept constant.
Both curves have similar shape but the difference the drifts
between a standard $\Lambda$CDM model and cosmological models
tends to zero at hight redshift. Fig.~\ref{AmpError} (right panel)
depicts this difference for two models with either $w = -0.95$ or
$w = -0.98$.

The feasibility of this measurement is most challenging and
impossible with present-day astronomical facilities. However, it
was  recently revisited~\cite{Loeb} in the context of the new
generation of Extremely Large Telescopes\footnote{ {\tt
http://www.eso.org/projects/e-elt/Publications/
ELT\_SWG\_apr30\_1.pdf}} (ELT), arguing that with such outstanding
collecting areas one could measure velocity shifts of order
$\delta v\sim 1-10\ {\rm cm/s}$ over a 10 year period from the
observation of the Lyman-alpha forest on QSO absorption spectra.
In particular, it is one of the main science driver to design the
COsmic Dynamics EXperiment (CODEX)
spectrograph~\cite{Pasquini1,Pasquini2} for the future European
ELT (E-ELT).

The performances of CODEX and its capability to measure a time
drift of very distant objects were estimated using Monte-Carlo
simulations of quasar absorption spectra. The expected velocity
accuracy of this experiment can be written as follows (see
Ref.~\cite{Pasquini1})
$$
 \sigma_v=1.4\left(\frac{S/N}{2350}\right)^{-1}
 \left(\frac{N_{\rm QSO}}{30}\right)^{-1/2}
 \left(\frac{1+z}{5}\right)^{-1.8}\,{\rm cm/s}\ ,
$$
provided the absorption lines are resolved. $S/N$ denotes the
signal-to-noise ratio, for a pixel scale of $0.0125\, {\rm\AA}$
and $N_{\rm QSO}$ is the number of quasars. Thus, spectroscopic
measurements of about 40 quasars with $S/N\sim2000$ ten years
apart can reach a $1.5\,{\rm cm/s}$ accuracy.  This is within the
reach of a CODEX instrument mounted on a $60-80$ meter ELT by
observing a 16.5$^{\rm th}$ magnitude QSO during  2000
hrs~\cite{Pasquini1}.

Many systematic effects that may spoil the time drift signal, such
as Earth rotation, proper motion of the source, relativistic
corrections etc., are discussed in Ref.~\cite{Pasquini1}. The
acceleration of the Sun in the Galaxy seems more a serious problem
because its amplitude  may be of the same order than the cosmic
signal. However, it has not been measured yet, so its nuisance is
still unknown. On the other hand, subtle contaminations like
accelerations produced by large scale structures have never been
estimated in the error budget. The purpose of this work is to
address this issue and to estimate whether it may hamper the
cosmological interpretation of the time drift.

\section{Cosmological perturbations}

Eq.~(\ref{SLformula}) relates the time drift of the observed
redshift to the Hubble function, assuming a perfectly homogeneous
and isotropic Friedmann-Lema\^{\i}tre spacetime. In the real
universe, however, velocity terms arising from cosmological
perturbations add up  as noise contributions and increase the
scatter of the redshift drift around its mean value.

The distribution of the redshift drift can be predicted using
the expression of $\dot z$ to first order in the cosmological
 perturbations. It is derived in the Appendix A of this work.
 At first order in the metric perturbations and in $v/c$ it writes
\begin{equation}
 \dot z=\dot{\bar
z}(\eta_0,z) + \zeta(\bx_\obs,\eta_0,\bm{e};z)  \ ,
\label{eq2}
\end{equation}
with
\begin{eqnarray}
 \zeta(\bx_0,\eta_0,\bm{e};z)
 & = & -\Phi_\obs\dot{\bar z}(\eta_0,z)+(1+z)\left[\bm{e}.\dot{\bm{v}}-\dot\Psi
 \right]^\obsup_\emit \!\!\!\!\!.
\label{eq3}
\end{eqnarray}
This formula involves both Bardeen potentials, $\Phi$ and $\Psi$,
and the peculiar acceleration, $\dot v$. A dot denotes a
derivative with respect to observer proper time and $\bm{e}$ is
the direction of observation. $O$ and $E$ refer to the observer
and emitter respectively (see the Appendix more precise
definitions of all the variables involved in this equation).

The first term at the right hand side of Eq.~(\ref{eq3}) clearly
arises from the local position of the observer. The second term of
Eq.~(\ref{eq3}) encodes Doppler effect due to the relative motion
of the observer and the source as well as the equivalent of the
integrated Sachs-Wolfe term~\cite{SW,pubook}. Eq.~(\ref{eq3})  is
the analog of the (direction dependent) temperature anisotropy of
the cosmic microwave background (CMB) compared to the mean CMB
temperature.

\section{Estimate of the variance}

The variance of  $\zeta(\bm{e})$ can be split into contributions
coming from the time dependence of the gravitational potentials,
$\zeta_{\dot\Phi}\equiv(1+z)\left[\dot\Psi\right]^\obs_\emit$, and
from the peculiar acceleration, $\zeta_{\dot
v}=(1+z)\left[\bm{e}.\dot{\bm{v}}\right]^\obs_\emit$.

The estimation of $\zeta_{\dot\Phi}$ demands a full description of
the time evolution of the potential, both at emission and
observing times. In the following we derive it and discuss its
properties using the linear cosmological perturbation theory. The
validity of this approach will be more thoroughly addressed in the
next section.

Using the linear theory of structure growth, the density contrast can be
split as $\delta= D(t ) \varepsilon({\bf x})$, where
$\varepsilon({\bf x})$ comprises all details on the initial conditions. The growth
rate $D_+$ is the growing solution of the equation
\begin{equation}\label{Dequation}
\ddot D(t)+2H\dot D(t)=\frac{3}{2}H^2\Omega_\mat(t)D(t),
\end{equation}
where $\Omega_\mat(t)$ is the time dependent reduced density
parameter for the gravitating matter (see ref.~\cite{revue} for
details). On sub-Hubble scales, Einstein equations imply that
$\Psi=\Phi$ and $\Delta\Phi=\frac{3}{2}H^2 \Omega_\mat\,a^2\,
\delta$.

As the redshift increases, the dynamics of the universe is closer
and closer to the one an Einstein-de Sitter Universe. $\Phi$ is
therefore almost constant and $\dot\Phi$ is expected to vanish.
This is no longer true at low redshift, when the cosmological
constant (or the spatial curvature) starts to dominate. Instead,
the time evolution of the potential writes
\begin{equation}\label{phidotphirelation}
\dot\Phi=H \Phi \left[f(t)-1\right] \ ,
\end{equation}
where $f(t)=\dd\ln D_+/\dd \ln a$ comprises the intrinsic
evolution of the potential produced by the growing perturbations.
In a flat $\Lambda$CDM, $f$ is explicitly given by
\begin{equation}
\!f(t)=\!1-\!\frac{6}{11}\frac{
_2F_1\left[2,\frac{4}{3};\frac{17}{6};-\sinh ^2\left(\frac{3
   \alpha  t}{2}\right)\right] \sinh ^2\left(\frac{3 \alpha  t}{2}\right)}
   { _2F_1\left[1,\frac{1}{3};\frac{11}{6};-\sinh ^2\left(\frac{3 \alpha
   t}{2}\right)\right]}
\end{equation}
where $\alpha\equiv H_0\sqrt{\Omega_{\Lambda0}}$ and where $_2F_1$
is a hypergeometric function.

Using Eq. (\ref{phidotphirelation}) it is then easy to express the
r.m.s. of $\dot\Phi$ from the r.m.s. of the mass density
fluctuations, $\sigma_\delta$, as derived from  the Poisson
equation. More precisely
\begin{equation}
 \sigma_\delta=\left[\int{\dd^3\bk\over(2\pi)^3}\,P_\delta(k)\right]^{1/2}\ ,
\end{equation}
where $P_\delta$ is the power spectrum of the density contrast,
$\langle\delta_{\bk}\delta_{\bk'}\rangle = P_\delta(k)
\Dirac(\bk+\bk')$, and  $\delta_\bk$ are the Fourier modes of
$\delta$. To estimate $P_\delta$, we adopt the prescription by
Bond {\it et al.}~\cite{bbks} for the transfer function and the
normalization $\sigma_{8}=1$. The redshift dependence of the power
spectrum is then the one of the growing mode, $D_+(z)$, normalized
to unity at $z=0$.

Turning to the gravitational potential, it appears that, in the
standard model of cosmology with a primordial spectrum
 of index $n_s\sim0.95$, the amplitude of the potential
fluctuations is IR divergent. However, since the previous
calculation is only valid  for sub-horizon modes, it is necessary
to introduce a cut-off for modes typically beyond  the Hubble
scale. The expected potential fluctuations then drop to more
realistic amplitudes of $\sigma_{\Phi}\simeq5\times10^{-5}$. It
follows that, for a source at redshift $z$, the r.m.s. of $\dot z$
induced by the time variation of the gravitational potential is
\begin{equation}
 \langle\zeta_{\dot\Phi}^2\rangle^{1/2}\left(z\right) =
 \frac{3}{2}(1+z)\Omega_{\mat
 0}  \ \left[f(0)-1\right]\sigma_{\Phi}\ ,
\label{zetadotphi}
\end{equation}
which is of order $\zeta_{\dot\Phi}\sim (1+z) \times 10^{-5} H_{0}$, a
small number indeed.\\

\begin{widetext}

\begin{figure}[htb]
 \centerline{\epsfig {figure=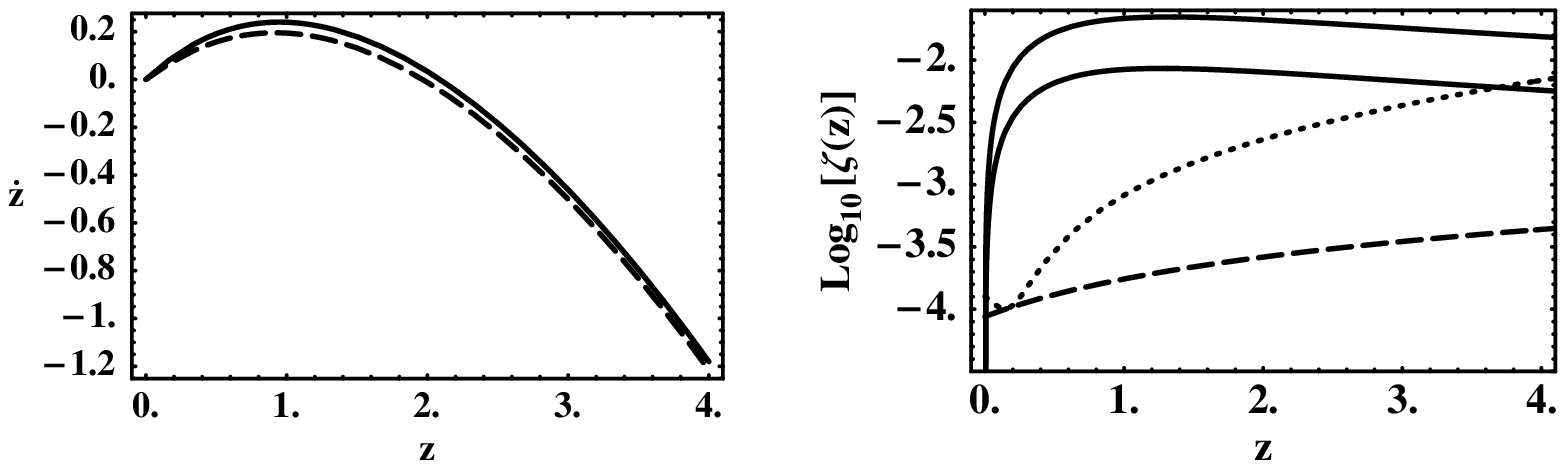,width=16cm}}
\caption{(left) The time drift of the redshift as a function of
the redshift of the source obtained from Eq.~(\ref{SLformula}) for
a $\Lambda$CDM model (solid line) and a model with a constant
equation of state $w = -0.9$ for the dark energy (dashed line).
(right) Amplitude of the r.m.s. of the systematic errors
$\zeta_{\dot v}$ due to cosmic acceleration effects. The
contribution of $\zeta_\obs$ (dashed line) is subdominant compared
to the one of $\zeta_\emit$ (dotted line). The solid lines
represents the difference between a standard $\Lambda$CDM model
and cosmological models with either $w = -0.95$ (upper solid line)
or $w = -0.98$ (lower solid line).}
 \label{AmpError}
\end{figure}
\end{widetext}

The contribution of the peculiar acceleration, $\zeta_{\dot v}$,
is less obvious to derive because we do not have a complete theory
that describes the expected distribution of the local
line-of-sight acceleration. However, in the cosmological linear
theory not only are the metric components supposed to be small (as
explicitly used above), but also the density contrast and the
velocity gradients (compared respectively to unity and $H$), see
Ref.~\cite{revue}.

The Lyman-alpha forest is believed to be dominated by low density
clouds of intergalactic medium, with individual accelerations
primarily triggered by large-scale structures. Assuming then
linear theory holds in our context, the local acceleration writes
$\dot{v}_{i}=-H v_{i}-\partial_{i}\Phi/a$, so that
\begin{equation}\label{zetaac2}
\zeta_{\dot v}(\bm{e},z)=(1+z)\  e^{i}\left[H(z)
v_{i}+\frac{1}{a}\partial_{i}\Phi\right]^\obsup_\emit\ .
\end{equation}
In terms of the dimensionless divergence $\theta(\bx)=\partial_{i}
v_{i}/aH$, the linear continuity equation reduces to
$\theta(t,\bx) = -f(t)\ \delta(\bx)$ at linear order. This implies
that the Fourier components of the velocity, density contrast and
potential are related by $k^2 H v_{i}(\bk)= f(t) a H^2
k_i\delta_\bk$ and $k^2\Phi_{,i}(\bk)/a = 3\Omega_{\mat} a\,H^2
k_{i} \delta_\bk/2$. Using our previous estimate of $P_\delta$,
one easily derives the r.m.s. of the two contributions to
$\zeta_{\dot v}$,
\begin{eqnarray}\label{zetaO}
 \langle\zeta_\obs^2\rangle^{1/2} &=&
      (1+z)\left[\frac{3}{2}\Omega_{\mat
      0}-f(0)\right]H_0^2 \ \hat\sigma\,
\end{eqnarray}
that depends on the emission time only through the factor $(1+z)$,
and
\begin{eqnarray}\label{zetaE}
 \langle\zeta_\emit^2\rangle^{1/2} &=&
 \left[\frac{3}{2}\Omega_{\mat}(t)-f(t)\right]H^2(t)D_{+}(t)\ \hat\sigma
\end{eqnarray}
where $\hat\sigma^2\equiv\displaystyle{\int\frac{\dd^3\bk}{(2\pi)^3}\frac{1}{3
k^2}P(k,z=0)}$.

These two terms are independent and should be summed
quadratically. The resulting r.m.s. of $\zeta_{\dot v}$ depicted
on Fig.~\ref{AmpError} (right panel)  shows  $\zeta_\emit$ is the
dominant contribution at all redshifts. It rises to a percent
level from $z=0$ to $z=4$. At $z\sim4$, $\zeta_{\dot v} \sim 0.5\%
$, while $\zeta_{\dot v}(\bm{e},z)$ is ten times smaller. Both
terms have similar behaviour and are basically unchanged for any
realistic flat cosmology having and effective $w$ close to $-1$,
but note that this is {\em a priori} not the case for any model.

\section{Discussion}

Assuming the cosmological time drift  derived from QSO absorption
lines by the Lyman-alpha forest may be contaminated by
extra-acceleration of clouds by massive structures, it is
legitimate to question the validity of the linear regime
approximation used throughout this work.

Let us first consider the acceleration of an absorbing Lyman-alpha
cloud. On large scales, clouds are located inside filaments
infalling towards massive clusters or super-clusters of galaxies.
Assume, then, the acceleration is due to the gravitational
attraction of a super-cluster with  typical mass of order
$10^{15}M_\odot$, localized at 10~Mpc from the cloud. The
Newtonian acceleration  is about $a_{\rm
N}\sim1.45\times10^{-15}\,{\rm km}/{\rm s}^2$. In comparison, the
Hubble acceleration $c H_0$  is $a_{\rm
H}\sim6.8\times10^{-13}\,{\rm km}/{\rm s}^2$ so that $a_{\rm
N}/a_{\rm H}\sim 2\times10^{-3}$.  This ratio may change by one
order of magnitude, depending on the mass and length scales one
may consider for clusters, super-clusters or  filaments, but is
always sufficiently small to keep the linear approximation valid.
It is also worth noticing its amplitude is close to theoretical
expectations derived in the previous Section. We therefore
speculate the simple interpretation of our theoretical estimate as
being primarily due to the acceleration of the nearest rich
cluster is pertinent\footnote{Liske {\em et al.} (in preparation)
also estimated the contamination of the drift signal produced by
peculiar motions. In contrast to our analysis done in a full
General Relativity context, they simply used Special Relativity
formalism. Note that \cite{peculaccel} derived the peculiar
acceleration of strong gravitational potentials like clusters of
galaxies but to predict the peculiar velocity drift over several
decades produced by nearby systems on a test particle. Both
results agree with our predictions.}. To confirm this and get more
sophisticated description of accelerations the
use of numerical simulations is indeed necessary.\\

In practice, a time drift is not measured from a single absorption
line but by averaging several lines spread over a spectral range
$\Delta \lambda$ defined by the spectrograph. If the acceleration
of Lyman-alpha clouds is primarily driven by clusters of galaxies
located around their neighborhood, then clouds are not dynamically
independent and accelerations of closeby clouds are correlated. We
are thus interested in the variance of $\dot z$, averaged over a
bound  comoving distance  $\Delta\chi$ along the line of sight,
$$
 \bar{\dot z} = \int_\chi^{\chi+\Delta\chi}\dot z(\chi')\dd\chi'\, .
$$
It is related to the variance from correlations obtained on a
single line by
$$
 \langle\bar{\dot z}^2\rangle = \alpha^2(\bar z,\Delta z) \langle
 \zeta^2(z)\rangle\ ,
$$
where the coefficient $\alpha(\bar z,\Delta z)$ depends on the
physical size over which the average is performed. $\Delta z$ is
the redshift range explored by the spectrograph at the mean
redshift $\bar z$: $\Delta z = (1+\bar z) \Delta\lambda/ \lambda$.
For a $\Lambda$CDM universe, it corresponds to a comoving distance
of $\Delta\chi=D_{H_0}[\Omega_{\mat0}(1+\bar
z)^3+\Omega_{\Lambda0}]^{-1/2}$ , with $D_{H_0}=3000h^{-1}$~Mpc.
$\alpha$ can be computed from the correlation of the acceleration
field,
$$
 \langle a(\chi_1)a(\chi_2)\rangle = \int \frac{\dd^3\bk}{(2\pi)^3}
 \hbox{e}^{i k_z(\chi_1-\chi_2)}\frac{P(k)}{3k^2}\ ,
$$
as
\begin{equation}
 \alpha^2 = \frac{1}{\hat\sigma^2}\int \frac{\dd k_z\dd^2\bk_\perp}{3 k^2}
 \frac{\sin k_z\Delta\chi}{k_z\Delta\chi} P(k) \ ,
\end{equation}
where $\Delta\chi$ is the size of the comoving radial distance over
which the average is performed.

\begin{figure}[htb]
 \centerline{\epsfig {figure=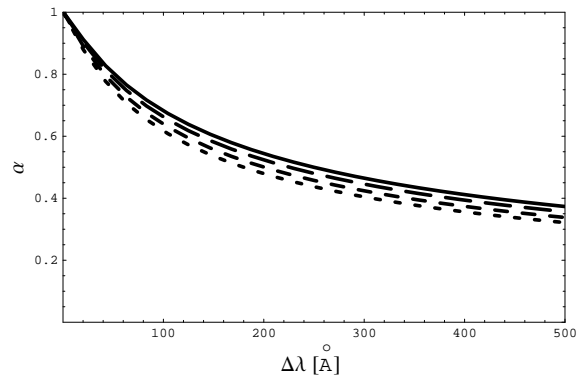,width=8cm}}
\caption{The coefficient $\alpha$ that enters in Eq.~(\ref{zf}) as
a function of the width of wavelengths, $\Delta\lambda$, on which
the observations are average for several source redshifts, $\bar
z=4$ (solid line), $\bar z=3$ (long dashed line), $\bar z=2$
(dashed line) and $\bar z=1$ (dotted line) for $\Lambda$CDM with
$h=0.7$, $\Omega_{\mat 0}=0.3$ and $\Omega_{\Lambda0}=0.7$.}
 \label{f2}
\end{figure}

If we could naively split the Lyman-alpha forest along a line if
sight into radial bunches of physically decoupled cloud systems,
without correlated accelerations, $1/\alpha^2$ would provide an
estimate of the  number of bunches. From an observational point of
view, $1/\alpha^2$  expresses the effective number of absorption
line systems without correlated acceleration probed by a
spectrograph covering a wavelength range $\Delta \lambda$ around
the mean redshift $\bar z$. It increases when the spectral
coverage of the spectrograph increases (see Fig. \ref{f2}). For
example, if $\bar z=4$, and  $\Delta\lambda= 100~{\rm\AA},
200~{\rm\AA}, 500~{\rm\AA}$, that is $\Delta z=0.1, 0.2, 0.5$ (see
Fig.~\ref{f2}) respectively, then $\alpha$ is 0.69, 0.55, and
0.38, and $1/\alpha^2=$ 2.1, 3.3 and 6.9. However, $1/\alpha^2$
only takes into account one light of sight. If we average over
$N_{\rm QSO}$ randomly selected lines of sight, then we expect
that
\begin{equation}\label{zf}
 \sigma_{\dot z} = \alpha(\Delta\lambda,\bar z)\,N_{\rm
 QSO}^{-1/2}\, \ \zeta(\bar z)\ .
\end{equation}
  Hence,  the spectral range of
the spectrograph together with the number of lines of sight can easily drop the contribution of
cosmological perturbation to the variance budget below a $0.1\%$ level.

It is interesting to notice that the theoretical values of
$1/\alpha^2$ derived in the previous paragraph can easily be
interpreted and predicted from simple physical arguments.  The
spectral range of a CODEX-like spectrograph as described in
Ref.~\cite{Pasquini1}, is $\Delta\lambda\sim 500{\rm\AA}$. At a
redshift of $\bar z\sim 4$, it corresponds to $\Delta z \sim
(1+\bar z)\Delta\lambda/ 5000{\rm\AA}\sim 0.5$ and to a comoving
length of $\Delta\chi\sim 300$~Mpc. If we assume that the
coherence scale of the velocity field is the typical size of the
super-cluster ($\sim $30~Mpc), then CODEX can probe about 10
independent systems per line of sight. This is of the same order
as $1/\alpha^2=6.9$ for $\Delta\lambda=500~{\rm\AA}$ discussed
above, which confirms its interpretation as an effective number of
uncorrelated cloud systems. It also simply explains why
$\sigma_{\dot z} \propto \alpha$, which is nothing but the inverse
square root of this number.

\section{Conclusion}

In order to measure the cosmological time drift of the redshift,
many systematic effects will have to be understood. Besides the
systematic errors of astrophysical origin that may affect the
observation of the Lyman-alpha forest, large scale structures will
induce a dispersion of $\dot z$. This work addresses this issue.

First, we have derived the expression of the time drift of
cosmological redshift at first order in the perturbation. This was
then used to estimate its variance and then to demonstrate it
depends on two  main effects, the accelerations and the local
gravitational potential at both the source and observer positions.

The contributions at the observer position have not been discussed
further. High precision astrometric observations with GAIA will
soon provide exquisite  knowledge of the motion of the Earth in
the Milky Way. It  will pinpoint all local acceleration terms with
enough accuracy to remove this contribution easily~\cite{gaiaref}.

In contrast, the contributions at the source position are much
more difficult to subtract. In the linear regime, we have shown
that the gravitational potential contribution is negligible while
the acceleration of the source is typically of the order of 1\% at
$z=2-4$. We argue a dominant contribution to this term is the
acceleration of galaxy clusters near the source.

In order to understand whether the amplitude of this variance can
be reduced,  we have estimated the effect of averaging the signal
over several absorption lines. One can either profit from the
total spectral range covered by the spectrograph  to measure the
drift from all lines detected along a line of sight, provided
correlated acceleration contributions are taken into account, or
use the mean  drift over many randomly selected lines of sight.
The first option reduces the variance by the square root of the
number of uncorrelated clouds systems along a line of sight, the
second by the square root of the number of independent lines of
sight. For an instrument having the current specifications of the
CODEX spectrograph, it is then easy to drop the contribution of
large-scale structures to the total variance of the time drift
down to a 0.1\% level.

\noindent{\bf Acknowledgements:} We thank Jochen Liske and Luca
Pasquini for providing Ref.~\cite{Pasquini1} before publication
and for their comments, Patrick Petitjean for useful discussions,
and Eric Linder his useful comments. We also thank St\'ephane
Charlot and Jean-Gabriel Cuby for organizing the Programme
National de Cosmologie discussion on ELT which triggered these
questions.


\pagebreak
\appendix
\section{Time drift at first order in the perturbation}

The redshift is defined as the ratio of the wavelengths measured
at the observer and the emission (galaxy) positions, both in their rest-frame. It can
be expressed in terms of the tangent timelike vector to the
observer (labelled $O$) and emitting galaxy (labelled $E$),
$u^\mu$, and the tangent vector $k^\mu$ to the null geodesic
joining $E$ to $O$ as
\begin{equation}
 1+ z = \frac{(u_\mu k^\mu)_\emit}{(u_\mu k^\mu)_\obs}\ .
\end{equation}
We want to express the redshift and its time drift at first
order in the perturbations around a Friedmann-Lema\^{\i}tre
spacetime with general metric $\dd s^2 = g_{\mu\nu} \dd x^\mu\dd
x^\nu = a^2\hat g_{\mu\nu} \dd x^\mu\dd x^\nu$ with
\begin{equation}
 \hat g_{\mu\nu} \dd x^\mu\dd x^\nu = -(1+2\Phi)\dd\eta^2
 +(\gamma_{ij}+h_{ij})\dd x^i\dd
 x^j\ ,
\end{equation}
where $h_{ij}=-2\Psi\gamma_{ij}$, i.e. we chose to work in
the Newtonian gauge and have neglected the effect of gravity
waves.

It is clear that if $k^\mu$ is the tangent vector to a null
geodesic of $g_{\mu\nu}$ then $\hat k^\mu=a^2k^\mu$ is the tangent
vector to a null geodesic of $\hat g_{\mu\nu}$. Decomposing $\hat
k^\mu$ as $\hat k^\mu= E(1+M, e^i+\delta e^i) \Longleftrightarrow
k^\mu= E a^{-2}(1+M, e^i+\delta e^i)$, where $E$ is a constant,
the geodesic equation reduces to (see e.g. Ref~\cite{ubcorde} for
details)
\begin{equation}\label{eq:A3}
 \frac{\dd M}{\dd\lambda} = -\Phi' -2e^i\partial_i \Phi -\frac{1}{2}h_{ij}'e^i e^j\ ,
\end{equation}
where $\lambda$ is an affine parameter along the null geodesic,
$\dd M/\dd\lambda \equiv M'+e^i\partial_i M$ and  a prime refers
to a derivative with respect to the conformal time.

At first order in the perturbations, the vector field $u^\mu$ is
explicitly given by $u_\mu =a(-1-\Phi,v_i)$. We then deduce
that $ k^\mu u_\mu=E\left[-1-M-\Phi +e^iv_i \right]/a$. Thus
\begin{equation}
 (1+z)= \frac{a(\eta_\obs)}{a(\eta_\emit)}\left\lbrace
 1 -[M+\Phi-e^iv_i]_\emit^\obsup
 \right\rbrace\ .
\end{equation}
Integrating Eq.~(\ref{eq:A3}), one derives that $[M]_\emit^\obsup=
-2[\Phi]_\emit^\obsup+\int_\emit^\obsup\left(\Phi'-\frac{1}{2}h_{ij}'e^i
e^j\right)\dd\lambda$. Therefore, the redshift writes
\begin{widetext}
\begin{equation}\label{egen}
 (1+z)= \frac{a(\eta_\obs)}{a(\eta_\emit)}\left\lbrace
 1 +[\Phi+e^iv_i]_\emit^\obsup - \int_\emit^\obsup
 \left(\Phi'+\Psi'\right)[\bm{x}(\eta),\eta]\dd\eta
 \right\rbrace
 \equiv
 \frac{a(\eta_\obs)}{a(\eta_\emit)}\left\lbrace
 1 +[\Upsilon]_\emit^\obsup
 \right\rbrace
 \ ,
\end{equation}
\end{widetext}
where we have shifted to the conformal time. This equation indeed
mimics exactly the standard Sachs-Wolfe formula~\cite{SW}.

At the background level,  the observer and emitter are comoving so
that their proper time corresponds to the cosmic time. It follows
that $\delta\eta_\emit=\delta\eta_0=\delta t_\obs/a_0$,  so
Eq.~(\ref{egen}) implies $\delta z=(1+z) (\mathcal{H}_0 -
\mathcal{H}_\emit)\delta\eta_0/a_0$, where $\mathcal{H}=a'/a$.
Shifting back to cosmic time, we get the standard expression for
the time drift  of a source located at redshift $z$, that is
Eq.~(\ref{SLformula}).

At first order, one has to take into account the motion of the
observer and the emitter, as well as the metric perturbations.
This will manifest in the difference between the cosmic time and
proper time.

If at the proper time $\tau_0$, the observer was located in
($\bm{x}_\obs,\eta_0$) and had a proper velocity $\bm{v}_\obs$,
then at a proper time $\tau_0+\delta\tau$
\begin{enumerate}
\item the cosmic time is
$\delta\tau = (1+\Phi_\obs)\delta t_0=
a_0(1+\Phi_\obs)\delta\eta_0$, up to terms in $v^2/c^2$ and
\item
the observer has moved to $\bm{x}'_\obs = \bm{x}_\obs +
\bm{v}_\obs\delta\eta_0$ so that he is located in
\begin{eqnarray}
 (\bm{x}_\obs',\eta_0') =
                       (\bm{x}_\obs,\eta_0) + (\bm{v}_\obs,1)\frac{1-\Phi_\obs}{a_0}\delta\tau\ .
\end{eqnarray}
\end{enumerate}
The null geodesic is still evaluated at the background level so
that $\bm{x}(\eta) =  \bm{x}'_\obs + \bm{e}(\eta'_0 - \eta)$,
where $\bm{e}$ is the direction of observation, and the emitter is
now located $\bm{x}'_\emit=\bm{x}_\emit +
\bm{v}_\emit\delta\eta_\emit$. This implies that
\begin{eqnarray}
 \delta\eta_\emit =\frac{1+\bm{e}.\bm{v}_\emit}{1+\bm{e}.\bm{v}_\obs}\delta\eta_0
                    \simeq
                    [1+\bm{e}.(\bm{v}_\emit-\bm{v}_\obs)]\delta\eta_0\
                    .
\end{eqnarray}
Thus, plugging these new positions in Eq.~(\ref{egen}), we obtain
that $\dot z=\dot{\bar z}(\eta_0,z) +
\zeta(\bx_\obs,\eta_0,\bm{e};z)$ with
\begin{eqnarray}
 \zeta(\bx_0,\eta_0,\bm{e};z)
 & = & -\Phi_\obs\dot{\bar z}(\eta_0,z)+(1+z)\left[\bm{e}.\dot{\bm{v}}-\dot\Psi
 \right]^\obs_\emit\ ,
\end{eqnarray}
where a dot refers to a derivative with respect to observer proper
time. This expression gives the full redshift drift at first order
in the metric perturbations and in $v/c$.

\end{document}